 \documentclass[preprint,5p,times,twocolumn,sort&compress]{elsarticle}

\usepackage{amssymb}
\usepackage{bm}

\usepackage[utf8]{inputenc} 
\usepackage{booktabs,amsmath}
\usepackage{bm,amsmath,amssymb,amsfonts,latexsym,psfrag}
\usepackage{amssymb}
 \usepackage{array}
\usepackage{mathtools}
\usepackage{amsthm}
\usepackage{newlfont}
\usepackage{esdiff}
\usepackage{multirow}
\usepackage{graphics}
\usepackage{graphicx}
\usepackage[normalem]{ulem}
\usepackage{color}
\usepackage{tabularx}
\usepackage{bm}
\usepackage{natbib}
\usepackage{url}
\usepackage{bm}
\usepackage[colorlinks=true,citecolor=cyan,urlcolor=blue,bookmarks=true,bookmarks=true,bookmarksopen=true,bookmarksnumbered=true,bookmarksopenlevel=3]{hyperref}

\definecolor{airforceblue}{rgb}{0.36, 0.54, 0.66}
\definecolor{steelblue}{rgb}{0.27, 0.51, 0.71}
\definecolor{amber}{rgb}{1.0, 0.49, 0.0}

\newcommand{\GB}{\text{\tiny GB}}
\newcommand{\DS}{\text{\tiny DS}}
\newcommand{\SM}{\text{\tiny SM}}
\newcommand{\DM}{\text{\tiny DM}}
\newcommand{\port}{\text{\tiny portal}}
\newcommand{\UG}{\text{\tiny UG}}
\newcommand{\eq}{{\textrm{eq}}}

\newcommand{\LamDS}{\Lambda_\DS}

\newcommand{\mpl}{m_{\text{\tiny pl}}}
\newcommand{\TDS}{T_\DS}
\newcommand{\TSM}{T_\SM}
\newcommand{\TDSc}{T_\DS^{\textrm{c}}}

\newcommand{\TDSfo}{\TDS^{\textrm{\tiny fo}}}
\newcommand{\cW}{c_{\textrm{\tiny W}}}

\newcommand{\GaGB}{\Gamma_\GB}

\newcommand{\vrel}{v_\textrm{rel}}
\newcommand{\nS}{n_{0^{++}}}
\newcommand{\nSS}{n_{2^{++}}}
\newcommand{\nVV}{n_{1^{+-}}}
\newcommand{\eDS}{e_\DS}
\newcommand{\eSM}{e_\SM}

\allowdisplaybreaks

\journal{Physics Letters B}

\begin{document}

\begin{frontmatter}



\title{ {\bf Smooth reheating and dark matter via non-Abelian gauge theory}
\tnoteref{doi_1}
}

\author[label1]{{\scshape S.~Biondini}}
\address[label1]{Department of Physics, University of Basel, Klingelbergstr. 82, CH-4056 Basel, Switzerland }
\ead{simone.biondini@unibas.ch}

\author[label2]{{\scshape H.~Kole\v{s}ov\'{a}}}
\address[label2]{Department of Mathematics and Physics, University of Stavanger,
4036 Stavanger, Norway}
\ead{helena.kolesova@uis.no}

\author[label3]{{\scshape S.~Procacci}}
\address[label3]{Departement of Theoretical Physics, University of Geneva, Quai Ernest Ansermet 24, CH-1211 Geneva, Switzerland}
\ead{simona.procacci@unige.ch}


\begin{abstract}

We demonstrate how a dark sector can provide a crucial link between a vacuum-dominated early universe and the later radiation-dominated epoch.
As an example, we consider a feeble axion-like coupling of 
dark non-Abelian gauge fields with
a single inflaton. 
In this scenario, a dark heat bath emerges at the end of inflation.
As the dark sector cools down, its gauge fields confine into composite glueball states.
The reheating of the visible sector is realized through portal interactions. 
After confinement, discrete symmetries 
protect part of the glueball states
that 
form relic dark matter,  
while the others keep decaying into Standard Model degrees of freedom.
The dark relic abundance depends mostly on the
departure from equilibrium 
dynamics after the phase transition, constraining the confinement scale of the dark sector.
Moreover, indirect detection and Big-Bang nucleosynthesis set bounds on the scale of portal interactions. 
We show that in a narrow but non-vanishing region of the parameter space, the correct dark matter abundance and sufficient reheating temperature are simultaneously reached.

\end{abstract}


\end{frontmatter}
\makeatletter
\def\ps@pprintTitle{%
   \let\@oddhead\@empty
   \let\@evenhead\@empty
   \let\@oddfoot\@empty
   \let\@evenfoot\@oddfoot
}
\makeatother

\section{Introduction}
\label{sec:introduction}
Understanding the universe in terms of a few fundamental fields is perhaps the holy grail of particle cosmology. Assuming inflation to explain the origin of primordial perturbations implies the existence of a transition period, when 
the cold and inflated universe starts being populated by particles originating from the conversion of the inflaton energy.  A successful nucleosynthesis demands for the onset of a Standard Model (SM) thermal bath at temperatures $\TSM \gtrsim 4 $ MeV \cite{Kawasaki:2000en,Hannestad:2004px,DeBernardis:2008zz}.  
Certainly, some mechanism had to be active and efficient in producing a cosmological abundance of dark matter, which is about five times larger than that of the ordinary baryonic matter \cite{Planck:2018nkj}.  Current precision measurements efficiently constrain inflationary models \cite{Planck:2018jri,BICEP2:2018kqh,BICEP:2021xfz}, as well as dark sectors \cite{Harris:2014hga,Elor:2015bho,Arcadi:2017kky}.

It is then quite appealing and nowadays accepted as part of the speculations in particle cosmology, that a dark sector may be coupled to the inflaton. This entails the possibility of a dark sector being populated first from inflaton decays, and the SM particles being in turn generated via portal interactions with the dark sector, see e.g.~\cite{Biondini:2020xcj,Berlin:2016vnh,Tenkanen:2016jic,Paul:2018njm}. 
The main scope of our work is to link the various stages of the cosmological evolution, \emph{from} inflation \emph{to} the onset of the SM nucleosynthesis, for a quite minimal hidden sector: a non-Abelian pure Yang-Mills (YM) theory \cite{Carlson:1992fn,Faraggi:2000pv,Boddy:2014yra,Soni:2016gzf,Kribs:2016cew,Halverson:2016nfq,Forestell:2016qhc,Forestell:2017wov,Carenza:2022pjd,Carenza:2023eua,Gross:2020zam,Redi:2020ffc,Yamada:2023thl}. 
Dreaming of a complete framework, here we want to keep track of the continuous transition from the vacuum-dominated (inflation) to the radiation-dominated (hot big bang) epoch. 
In this work, we restrict to an inflaton solely coupled to a dark SU(N) gauge theory. 

As far as the inflationary stage is concerned, we consider 
axion-like (or natural) inflation \cite{Freese:1990rb}  as the minimal inflation model, where a weakly-coupled thermal bath does not imply large back-reactions to the slow-roll regime \cite{Berghaus:2019whh,Laine:2021ego,Klose:2022knn,Klose:2022rxh}. The setup for the evolution equations that connect the inflaton field with the non-Abelian sector, together with a detailed study of the temperature of the YM plasma, have been put forward in ref.~\cite{Kolesova:2023yfp}.  We follow up on the same scenario of a warm end of inflation by investigating the fate of the dark sector, 
asking whether the observed dark matter energy density \emph{and} the generation of a sufficiently hot SM bath could be realized. 

The salient features of the dark sector are dictated by the behaviour of the non-Abelian gauge group. As 
the gauge coupling becomes
large at low energies, 
dimensional transmutation
implies 
a scale similar to 
$\Lambda_{\textrm{\tiny QCD}}$ for the SM 
strong interactions.  
We refer to this scale as $\LamDS$ in the following. 
During the universe's thermal history, whenever the temperature of the dark sector crosses the scale $\LamDS$ (from above), the physical
degrees of freedom become a tower of hidden glueballs, 
as a result of a strongly-coupled dynamics. 

Glueball states are often classified in terms of their quantum numbers under angular momentum ($J$), parity ($P$), and charge
conjugation ($C$), which ultimately determine the effective interactions with the visible sector~\cite{Juknevich:2009gg,Juknevich:2009ji}. Glueballs can then generically decay into SM particles and provide a way to reheat the SM. However, some of the glueball states can be rendered cosmologically stable. More specifically, C-odd glueballs feature a suppressed decay rate with respect to C-even states, hence making the lightest C-odd glueball a viable dark matter candidate \cite{Forestell:2017wov}. This is the option that will be explored in our work.

We show that it is challenging but possible to account for the observed dark matter relic density \emph{and} to reheat the SM at temperatures larger than 4 MeV, while keeping the C-odd glueball dark matter cosmologically stable. Our study is complementary to earlier investigations \cite{Forestell:2016qhc,Forestell:2017wov,Gross:2020zam,Redi:2020ffc,Li:2021fao}, where the dark non-Abelian plasma is absent or very sub-dominant during the early stages of the universe thermal history, namely the inflaton is assumed to preferably reheat the SM. 
 
The structure of the paper is as follows. In section~\ref{sec:model} we discuss the model for inflation and the dark sector, together with the portals with the SM. We present the evolution equations for the various components in section~\ref{sec:ev_equations} and their numerical solution, together with the viable parameter space, in section~\ref{sec:results}. We offer some discussion and concluding remarks in section~\ref{sec_discussion}.

\section{Setup}
\label{sec:model}
This section introduces 
the benchmark models we use to describe inflation, the emergence of a dark sector, and the interactions of the latter with the SM.

\subsection{Minimal non-Abelian axion-like inflation}
We consider a variant of axion-like inflation~\cite{Freese:1990rb} where the inflaton couples feebly to a non-Abelian dark sector,
resulting in the 
production of YM fields
already during the slow-roll inflationary phase \cite{Berghaus:2019whh,Laine:2021ego,Kolesova:2023yfp}.
Denoting $\varphi$ the inflaton field, the
Lagrangian reads  
\begin{equation}
    \mathcal{L} \supset \frac{1}{2} \partial_\mu \varphi \, \partial^\mu \varphi - V(\varphi) - \frac{\varphi}{f_a} \, \frac{\alpha \, \epsilon^{\mu \nu \rho \sigma} \,  F^c_{\mu \nu} F^c_{\rho \sigma}}{16 \pi } + \mathcal{L}^{\textrm{dark}}_{\textrm{\tiny YM 
    }} \, .
    \label{eq:Linfl}
\end{equation}
The dark YM sector is described by the field strength, 
$F^c_{\mu \nu}$,
the coupling, $\alpha$, 
a colour index, $c\,{\in}\,\lbrace1,\ldots, N\rbrace$,
and $\mathcal{L}^{\textrm{dark}}_{\textrm{\tiny YM
}}$, which comprises the SU$(N)$ kinetic term. Notice that we set $N=3$ in this work. 
For quantitative studies, we set 
the inflaton potential $V(\varphi)= m_\varphi^2 f_a^2 [1 - \cos(\varphi/f_a)]$, 
and fix 
decay constant, $f_a$, and mass, $m_\varphi$, 
using Planck's constraints \cite{Planck:2018jri}, as in ref.~\cite{Klose:2022rxh}.\footnote{The values we use are $f_a \approx 1.25 \, \mpl$, $m_\varphi \approx 1.09 \times 10^{-6} \, \mpl$. 
This potential is known as {\it natural} 
since it might be generated in a UV completion of the theory. 
} 
However, our final results do not depend on the details of the slow-roll phase, 
as discussed in sec.~\ref{sec_discussion}. 

The heating-up dynamics of the system in eq.~\eqref{eq:Linfl} has been studied in ref.~\cite{Kolesova:2023yfp}, showing a crucial dependence on the assumed confinement scale of the dark non-Abelian bath, $\LamDS$. 
In particular, if $\LamDS$ is small with respect to the scale of inflation, the energy released by the weakly coupled inflaton suffices to maintain the bath at a temperature above the critical scale, $\TDSc \simeq 1.24\,\LamDS$ \cite{Francis:2015lha}. 
A phase transition must then occur in 
a later stage, while the bath is cooling down after inflation.
In the present work, we show that adding a SM bath to 
this scenario 
can lead to successful reheating, 
and provide a viable dark matter candidate. 
Portal interactions between the SM and the dark sector 
might be active in both 
stages: first, when the dark vectors are weakly coupled,  
and later, 
when they are confined in composite glueball states.

\subsection{Dark vectors}
In the deconfined phase, the non-Abelian dark sector is expected to thermalize\footnote{This was studied in the context of heavy-ion collisions, see e.g.~\cite{Fu:2021jhl}. We verify the assumption for our scenario a posteriori.} and for a given confinement scale $\LamDS$, its state can be described solely by 
its temperature, $\TDS$.
In order to capture the onset of non-perturbative dynamics approaching the critical point, we use a lattice-fit approach \cite{Kolesova:2023yfp,Giusti:2016iqr}.
Also for the SM we assume that it is in a thermal state within itself,\footnote{We assume a radiation equation of state in the evolution equations and solve for the energy density, $e_\textrm{\tiny SM}$, avoiding parameterizations in terms of $T_\textrm{\tiny SM}$.}
meaning that the dynamics of the system in this regime is governed by equilibrium attractors. 

The universality of gravitational interactions provides an unavoidable source for the visible sector from a dark thermal bath. 
SM particles are produced through gravitational interactions at tree level, through 
the exchange of a graviton in the s-channel \cite{Garny:2015sjg,Tang:2016vch}.\footnote{The original works focus on the reverse setting, i.e.~a dark sector that is produced from SM collisions as mediated by gravitons.} 
A second option, which is often considered in the literature, envisages massive mediator
states that couple to both the dark and visible sectors. If the characteristic mass
scale of the mediators satisfies $M \gg \LamDS, \TDS$,
an effective description of the interactions can be employed and
leading operators have
mass dimension six and eight \cite{Juknevich:2009gg,Juknevich:2009ji}. 

In particular, the 
dimension-six operator that involves the SM Higgs doublet, $H$, reads
\begin{equation}
\mathcal{L}_6^\port \supset c_6 \frac{\alpha \,  \textrm{tr}(F F) H^\dagger H}{M^2}\,,
\label{op6}
\end{equation}
where $c_6$ is the matching coefficient, 
while color and Lorentz indices are suppressed. 
Whenever $M \ll \mpl$ the corresponding effective operators 
are expected to be more efficient for the production of the visible sector through scatterings. 
We adopt $M$ as a free parameter of the model. 
The corresponding rates for the SM production are estimated as $\Gamma_{\mathcal{G}} \simeq n_{\mathcal{G}}^\eq \langle \sigma_{\mathcal{G} \mathcal{G} \to \textrm{SM} \textrm{SM} } \vrel \rangle$ with $n_{\mathcal{G}}^\eq$ being the equilibrium number density for the dark gluons. In particular,
\begin{align}
   & \Gamma^{\textrm{{grav}}}_{\mathcal{G}} = \frac{\TDS^5}{\mpl^4} \frac{283}{ \, 40 \pi^3 \zeta(3) } \, ,
   \label{NRO_grav_vectors}
   \\
   & \Gamma_{\mathcal{G}} =  \frac{\TDS^5}{M^4} \frac{12 \, \alpha^2 \,  c_6^2}{ \pi^3 \zeta(3)} \, .
   \label{NRO_vectors}
\end{align}
The graviton-mediated thermally averaged cross-section in eq.~\eqref{NRO_grav_vectors} is extracted from refs.~\cite{Tang:2017hvq,Gross:2020zam}, bearing in mind that one needs dark vectors annihilation into SM particles;  we carry out a similar calculation for the 
operator in \eqref{op6}.\footnote{In both cases of eqs.~\eqref{NRO_grav_vectors} and \eqref{NRO_vectors}, the thermally averaged cross-section can be carried out analytically, whenever a Maxwell Boltzmann distribution is assumed for the dark vectors. We made this approximation in our work.\label{ftn:thermalAverage}} 
The rate in eq.~\eqref{NRO_vectors} is the leading contribution for the broken and unbroken phases of the SM and it corresponds to the processes $\mathcal{G} \mathcal{G} \to h h$ and $\mathcal{G} \mathcal{G} \to H H^\dagger$ respectively.  In the broken phase, the $2{\to} 2$ processes that are mediated by a  Higgs-boson exchange, $\mathcal{G} \mathcal{G} \to h \to \textrm{SM} \, \overline{\textrm{SM}}$, are  suppressed as $ (v_h/ \TDS)^2 \ll 1$. Dimension-eight operators are instead suppressed by $(\TDS/M)^4$.

We verify a posteriori that the portal interactions are never efficient enough to set equal the temperatures in the visible and dark sectors.
The maximal SM temperature reached in the benchmark viable scenario illustrated in fig.~\ref{fig:energy_densities} is $T_\SM \approx 3.5\times 10^6\,$GeV$\,\ll T_\DS \approx 8.4\times 10^{10}\,$GeV.

\subsection{Dark glueballs}
\label{subsec:model_glue}
Relevant cosmological information is contained in the departure from equilibrium of the dark glueballs after the phase transition. 
The glueball spectrum for 
SU$(3)$ YM theories has been studied on the lattice \cite{Morningstar:1999rf,Chen:2005mg,Mathieu:2008me}. Since the minimal YM action respects angular momentum, charge, and parity, glueballs are systematized in terms of the relevant quantum numbers $J^{PC}$. 

We 
present our analysis 
illustrating the dynamics of the lightest $C$-even and 
$C$-odd glueball, with masses $m_{0^{++}}{\simeq}\, 6.8 \LamDS$ and $m_{1^{+-}} {\simeq}\, 11.5 \LamDS$, respectively.\footnote{More precisely, $m_{J^{PC}} = \kappa_{J^{PC}}/r_0 $, where $r_0 \LamDS = 0.62(2)$ for SU(3) \cite{Aoki:2013ldr} and the values $\kappa_{J^{PC}}$ are collected e.g. in~refs.~\cite{Morningstar:1999rf,Meyer:2008tr}.} 
This minimal system has been shown to capture qualitatively the dynamics of the glueball ensemble \cite{Forestell:2016qhc,Forestell:2017wov}. 
However, 
we need to model the thermodynamics of the dark glueball system also at temperatures close $T_\DS^c$, where 
other glueballs are considerably abundant. To verify the two-glueballs approximation, we thus add the next lightest 
glueball, $2^{++}$ with $m_{2^{++}} {\simeq}\, 9.4 \LamDS$, to our numerical analysis.

Dark glueballs interact with the SM particles via the same set of effective operators as dark vectors~\cite{Juknevich:2009gg,Juknevich:2009ji,Gross:2020zam}, originating from graviton-mediation or a new-physics scales. However, such operators are now projected onto the composite states, hence requiring non-perturbative inputs \cite{Morningstar:1999rf,Chen:2005mg,Mathieu:2008me}. 
Moreover, the qualitative difference between $C$-even and $C$-odd states 
shapes the phenomenology. 
The lightest C-even glueball can decay through the dimension-6 operator~\eqref{op6}, 
with the decay rate scaling as $\LamDS^5/M^4$. 
Going beyond the parametric estimation, we adapt the expression for the dominant decay process $0^{++} \to hh$ 
from refs.~\cite{Juknevich:2009gg,Forestell:2017wov} 
for $m_{0^{++}} \gg m_h$,\footnote{Similarly as 
in the deconfined phase, the dominant decay width corresponds to the 
$0^{++} {\to} hh$ process.  
Decays into other SM particles are largely suppressed by $(v_h/m_{0^{++}})^2 {\ll} 1$ in our parameter space of interest. 
This statement is independent of the new physics scale $M$.} 
\begin{equation}
    \Gamma_\GB \equiv \langle \Gamma_{0^{++}}^{}\rangle \simeq \frac{c_6^2}{32 \pi M^4 \, m_{0^{++}}} (\alpha \, F_S^{0^{++}})^2 \left.\frac{K_1(z)}{K_2(z)}\right|_{z=m_{0^{++}/T_{\scalebox{.45}{DS}}}}\, . 
\label{GammaGB}
\end{equation}
Although non-perturbative inputs for the matrix element exists also from lattice studies, $4 \pi \alpha F_S^{0^{++}} {=}\, 2.3(5) m_{0^{++}}^3$ \cite{Chen:2005mg,Meyer:2008tr}, we rather relay to the large N-limit to align eq.~\eqref{GammaGB} with the approximation in eq.~\eqref{GammaDM}, $F_S^{0^{++}} {\sim}\, m_{0^{++}}^3/\alpha$ \cite{Forestell:2017wov}. 
The thermal average is represented by the factor $K_1/K_2\,{\in}\, [0.8,1.0]$, with $K_n~=~z^{-n}\int_z^\infty \!\!\text{d} x\, x^{n-1} e^{-x} \sqrt{x^2-z^2}$ extracted as in footnote \ref{ftn:thermalAverage}. 

Conversely, the lightest C-odd state can only decay via dimension-8 operators of the schematic form \cite{Juknevich:2009gg}
\begin{equation}
\mathcal{L}_8^\port \supset c_8 \, \frac{ \alpha^{3/2} \, \alpha_\text{\tiny{QED}}^{1/2} \, \textrm{tr}(F F F)_{\mu \nu} B^{\mu\nu}}{M^4}\,,
\label{op8}
\end{equation}
where $B_{\mu\nu}$ is the SM hypercharge field strength and  $c_8$ is again a matching coefficient. This operator 
induces the decays $1^{+-} {\to}\, 0^{++} \gamma$ and $1^{+-} {\to}\, 0^{++} Z$, 
with widths scaling as $\LamDS^9/M^8$. 
The $1^{+-}$ state 
is cosmologically stable, if 
its lifetime exceeds the age of the universe, yielding a dark matter candidate, even if 
C-even glueballs decay before BBN. 
In the $m_{1^{+-}} ,m_{0^{++}} \gg m_Z$ approximation, the full decay width is
\begin{align}
    \Gamma_\DM^{} \equiv \langle \Gamma_{1^{+-}}^{}\rangle \simeq c_8^2 \frac{\alpha_{\text{\tiny QED}}}{24 \pi \, \cW^2} \frac{m_{1^{+-}}^3 \left( \alpha^{3/2} M^{1^{+-}}_{0^{++}}\right)^2}{M^8} \left(1 - \frac{m_{0^{++}}^2}{m^2_{1^{+-}}} \right)^{3} \, ,
    \label{GammaDM}
\end{align}
where $\cW$ is the cosine of Weinberg angle. 
We 
omit the thermal average suppression, 
since $K_1/K_2\,{\to}\, 1$ at the temperatures relevant for 
this decay. 
For the matrix element $M^{1^{+-}}_{0^{++}}$, 
lattice results are 
\emph{not} available. 
The large-$N$ limit suggests $M^{1^{+-}}_{0^{++}} \alpha^{3/2} {\sim}\,  \sqrt{4 \pi/N} \, m_{0^{++}}^3 $ \cite{Forestell:2017wov}.

In addition to glueball-SM interactions, the dark composite states will interact with one another as a remnant of the strong dynamics that bind them. The corresponding processes, most notably $3{\to} 2$ and $2{\to} 2$ reactions, are key to determining the cosmological evolution of their densities (see sec.~\ref{sec:ev_equations}). 
Despite the interactions between glueballs being hard to determine from first principles, one expects the multi-glueball processes to be consistent with 
$J$, $P$, and $C$ conservation, and we again borrow parametric estimates for 
interaction strengths from results in the 
large-$N$ expansion (see \ref{app1_Glueball_Lag} for details).

\section{Evolution equations}
\label{sec:ev_equations}
In order to track the evolution of the different species, we rely on a network of evolution equations, that comprise 
the interaction rates introduced in sec.~\ref{sec:model}, and the Hubble rate, 
\begin{equation}
    H^2= 
    \frac{8 \pi}{3 \mpl^2} \left( e_{\varphi}  + \eDS + \eSM \right) 
    \, .
    \label{Hubble}
\end{equation}
We use the approximate initial value 
$H_\text{\tiny ref}^2\,{\equiv}\, 8\pi V(\varphi_0)/(3 \mpl^2)$ as an inverse-time unit.\footnote{The slow-roll inflation dynamics is an attractor solution \cite{Liddle:1994dx}, the initial value $\varphi_0\approx 3.5 \mpl$ is thus unimportant.}
The energy density of inflaton, $e_\varphi$, dark sector, $e_\DS$, and SM, $e_\SM$, follow coupled evolution equations, which we now turn to study. A benchmark solution is shown in fig.~\ref{fig:energy_densities}. 
\subsection{Energy densities}
The evolution of the inflaton embedded within a medium is taken over from refs.~\cite{Kolesova:2023yfp,Laine:2021ego}. 
In the post-inflationary era, the inflaton energy density is dominated by the kinetic term, while its pressure vanishes.\footnote{In particular, when the inflaton field rapidly oscillates around the minimum of the potential, $p_\varphi = \dot{\varphi}^2/2 - V $ averages to zero and $\dot \varphi^2 = e_\varphi + p_\varphi$ in the evolution equations can be effectively replaced by the more convenient variable $e_\varphi$. 
}
If we consider also the energy transfer between dark and visible sectors, the evolution of the system can be described by the following set of coupled equations, 
\begin{align}
     \dot e_\varphi + 3H e_\varphi 
    &= - \Upsilon e_\varphi 
    \label{eq:EvInf_dec}\\
    \dot{e}_\DS + 3H (\eDS + p_\DS) 
    &= \Upsilon e_\varphi 
    - \Gamma 
    \, \eDS  
    \label{eq:EvDS_dec}
    \\
    \dot{e}_\SM + 4H \eSM  
    &= \Gamma 
    \, \eDS \, . \label{eq:SM_dec}
\end{align}
The friction coefficient $\Upsilon$ in eqs.~\eqref{eq:EvInf_dec} and \eqref{eq:EvDS_dec}, 
transfers energy from the inflaton to 
the dark sector degrees of freedom. 
For temperatures $T{\ll} m_\varphi$ relevant in our setup, 
it is dominated 
by the vacuum decay rate $\Upsilon \simeq \alpha^2 m_\varphi^3/(32\pi^3 f_a^2)$~\cite{Laine:2011xm,Laine:2021ego,Kolesova:2023yfp}. 

As long as  
the dark sector stays in the deconfined phase, 
we parameterize its  
pressure $p_\DS$ and energy density $\eDS$ by a  
temperature, $\TDS$, using SU$(3)$ lattice fits~\cite{Giusti:2016iqr}.
Even if the friction term is suppressed, $\Upsilon\,{\ll}\, H$, in this first stage, dark vectors are efficiently populated. 
Relying on their fast thermalization \cite{Fu:2021jhl,DeRocco:2021rzv}, the SM is in turn produced from frequent $2{\to} 2$ scatterings of vectors \emph{\`{a} la} freeze-in. The corresponding rate is  
$\Gamma = \Gamma_{\mathcal{G}}$ as given in eq.~\eqref{NRO_vectors}.  

During the phase transition at $\TDS=\TDS^c$, the dark sector is in a mixed state, where dark vectors and glueballs are present at the same time \cite{Kolesova:2023yfp}. 
In this regime, within the adiabatic approximation, we assume equilibrium number densities for the glueball states and take into account the different decay rates to SM for the two phases.

Once all vectors have been confined into composite states, 
they are expected to remain 
in thermal equilibrium until a certain freeze-out temperature $\TDSfo \,{<}\, \TDSc$. 
The thermodynamic quantities describing the full dark sector, $\eDS$ and $p_\DS$, are then 
functions of 
glueball number densities and $\TDS$, according to the non-relativistic relations\footnote{
For SU$(3)$ we have $m_{0^{++}}\sim 5.4 \TDSc$, hence the non-relativistic approximation 
is not appropriate shortly after the phase transition. However, we verify 
the negligible effect of this error 
on our conclusions, mainly because the freeze out of stable glueballs happens at lower temperatures $T/T_c \sim 0.5$, see Fig.~\ref{fig:GB_freeze_out}.}
\begin{equation}
    \eDS= \sum_{i
    } m_i n_i \left( 1+ \frac{3}{2}\frac{T_\DS}{m_i} \right) \, , \quad p_\DS = \TDS \sum_{i
    }  n_i \,  
    \label{glue_e_p}
\end{equation}
where $i$ runs over different glueball states.
The modeling in eq.~\eqref{glue_e_p} is supported by lattice simulation of a pure SU(3) gauge theory~\cite{Meyer:2009tq}.\footnote{More specifically, for $\TDS/\TDSc \lesssim 0.7$, the lattice results agree very well with a system modeled by the two lightest 
glueballs $0^{++}$ and $2^{++}$. }
Eqs.~\eqref{eq:EvInf_dec}--\eqref{eq:SM_dec} for the energy densities need hence to be complemented by Boltzmann equations for the glueball number densities, capturing their departure from equilibrium (c.f. sec.~\ref{subsec:freezout}). 

In the confined phase, the inflaton can still lose energy in the form of dark vectors, which hadronize quickly and effectively contribute to the glueball energy density. 
However, in our parameter space of interest, the inflaton field usually vanishes 
around the time of the phase transition and does not affect the freeze-out dynamics.

At the same time, C-even glueballs decay into SM particles, hence $\Gamma\,e_\DS$ in eqs.~\eqref{eq:EvDS_dec} and~\eqref{eq:SM_dec} has to be replaced by $\GaGB e_\UG$, where $\GaGB$ was introduced in eq.~\eqref{GammaGB} and $e_\UG$ is the energy density in unstable glueballs. 
In turn, C-odd glueballs remain cosmologically stable.

\subsection{
Boltzmann equations for dark glueball states}
\label{subsec:freezout}

This is how we model the freeze-out dynamics of the glueballs, that 
eventually sets the dark matter relic abundance. 

As the dark sector is made of glueballs with different masses and interactions, the composite states fall out of chemical equilibrium at different temperatures and develop individual chemical potentials.
In order to track the glueball number densities, we employ a network of Boltzmann equations that is based on earlier studies~\cite{Forestell:2016qhc}, augmented by the feed-down of glueballs from the decaying inflaton, as well as the glueballs decay into SM.
It was shown in~\cite{Forestell:2016qhc} that a simplified scenario for the densities of the $0^{++}$ and $1^{+-}$ glueballs captures to a very good approximation the freeze-out dynamics of the whole ensemble. For simplicity, we adopt this choice when presenting the coupled Boltzmann equations in the main body of the paper. Hence, $e_\UG \simeq e_{0^{++}}$ and the stable glueball energy density is stored in $e_{1^{+-}}$.
We stress that, in order to improve the thermodynamic description of the YM sector~\cite{Meyer:2009tq},
we shall extend the treatment by including the next-to-lightest glueball state $2^{++}$ in our numerical analysis (see section~\ref{sec:results}). We collect the corresponding equations in \ref{app2_Glueball_3}.

The coupled Boltzmann equations for the two-glueballs system read 
\begin{align}
    \dot{n}_{1^{+-}}  + 3H \nVV 
    =&\, B_{\varphi}^{1^{+-}} 
    \Upsilon\frac{e_\varphi}{m_{{1^{+-}}}}
    -\langle \sigma_{2 {\to} 2} v \rangle \left[ \nVV^2 - \left( \frac{\nS}{\nS^\eq} \right)^2  (\nVV^\eq)^2 \right]  \,,\label{eq:Ev_6_NR_FO} \\
    \dot{n}_{0^{++}} + 3H \nS   
    =&\, 
    B_{\varphi}^{0^{++}} 
    \Upsilon\frac{e_\varphi}{m_{{0^{++}}}}
    +\langle \sigma_{2 {\to} 2} v \rangle \left[ \nVV^2 - \left( \frac{\nS}{\nS^\eq} \right)^2  (\nVV^\eq)^2 \right]
   \nonumber \\
    &- \nS \Gamma_\GB
    -  \langle \sigma_{3 \to 2} v^2 \rangle \,\nS^2(\nS- \nS^\eq )\, .\label{eq:Ev_1_NR_FO} 
\end{align} 
Thermally averaged cross-sections $\langle \sigma_{2 {\to} 2} v \rangle$ (here for 
$1^{+-}  1^{+-} {\to}\, 0^{++} 0^{++}$) and $\langle \sigma_{3 {\to} 2} v^2 \rangle$ (here 
changing the number of $0^{++}$ states) are extracted from ref.~\cite{Forestell:2016qhc}, as given in~\ref{app1_Glueball_Lag}. 

The relative abundance of different glueball species 
sourced (non-thermally) from the inflaton decays may be inferred phenomenologically from statistical hadronization models. Most notably, the branching ratio of a given glueball is proportional to its equilibrium distribution evaluated at a \emph{hadronization temperature}, $T_{\textrm{had}} \simeq \TDSc$ \cite{Andronic:2017pug,Batz:2023zef}. This is how we extract the branching ratios $B_{\varphi}^{\scalebox{.55}{$J^{PC}$}}$ %
that enter eqs.~\eqref{eq:Ev_6_NR_FO} and \eqref{eq:Ev_1_NR_FO}. 
These terms are of little importance below a certain threshold value of 
$\LamDS$,\footnote{ To estimate this threshold value $\LamDS^\varphi$, 
we observe that, when the inflaton decays, the Hubble rate satisfies $H\sim \Upsilon$. 
Moreover, the energy density of the dark sector around 
the phase transition can be approximated as $e_{\DS} \sim 0.1 \LamDS^4$. If we assume that $e_{\DS}$ 
dominates the Hubble rate we find 
$    \LamDS^\varphi/\mpl \sim \sqrt{\Upsilon/\mpl}~{\simeq}~10^{-12}.
    \label{lamphi}$} 
since then the phase transition in the dark sector 
happens after $e_\varphi{\to}\,0$.

\begin{center}
\begin{figure}
    \centering
    \includegraphics[width=.9\linewidth]{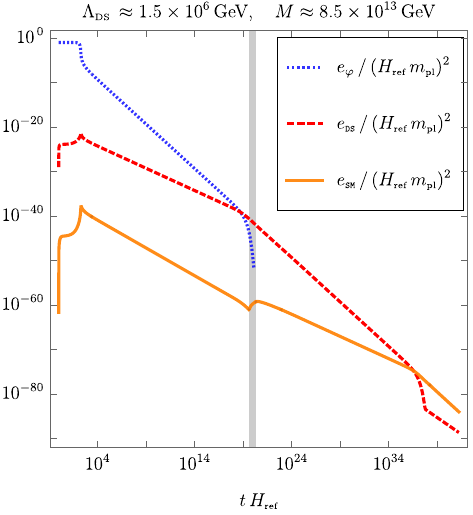}
    \caption{Numerical solution for the energy density for the various components. The dotted line stands for the inflaton, the dashed line for the dark YM, and the solid curve for the SM. The vertical gray band indicates the time at which     occurrence of the phase transition in the YM sector.}
    \label{fig:energy_densities}
\end{figure}
\end{center}

\section{Numerical solutions and constraints}
\label{sec:results}
\begin{figure}[t!]
    \centering
    \includegraphics[width=.9\linewidth]{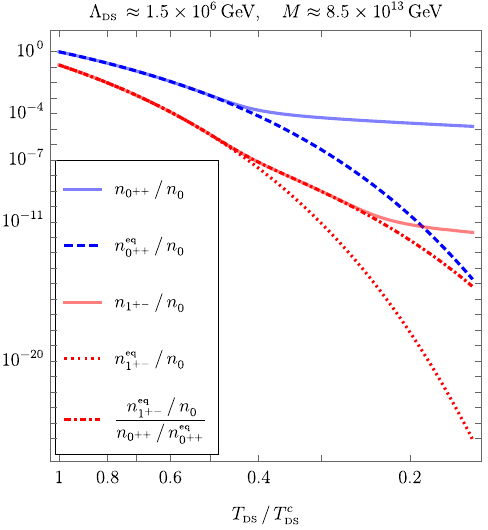}
    \caption{Number densities for the C-even $0^{++}$ and C-odd $1^{+-}$ glueballs respectively in blue (upper) and red (lower) solid lines. 
   By $n_i^\text{eq}$ we denote the equilibrium density with zero chemical potential. The normalization $n_0^{}\,{\equiv}\, \nS ( \TDS^c)$ corresponds to the number density of $0^{++}$ at the end of the phase transition.}
    \label{fig:GB_freeze_out}
\end{figure}
The main result 
of our study is a scan of numerical solutions for different values of the parameters $\LamDS$ and $M$.
An example of full solution for the energy densities of different sectors throughout the different stages described in sec.~\ref{sec:ev_equations} is shown in fig.~\ref{fig:energy_densities}.\footnote{For the parameters choice $\LamDS\approx 1.5 \times 10^6\,$GeV, and $M\approx 8.5\times 10^{13} \,\text{GeV}$ we obtain $T_\SM^\text{rh}\approx 4.3\,$MeV, and $\Omega_\DM \, h^2\approx 0.11$.} 
We rescale dimensionful quantities by powers of $H_\text{\tiny ref}$, defined below eq.~\eqref{Hubble}.

The period when $e_\varphi\approx \,$const 
corresponds to the slow-roll phase of inflation. 
When $\varphi$ approaches the minimum of the potential, $e_\varphi$ starts decreasing and the expansion of the universe becomes decelerated, translating in a milder dilution of $e_\DS$ and $e_\SM$. 
At the same time, both dark and visible sectors are still sourced by the highly oscillating inflaton, so that, at this stage, they reach their maximum temperatures. 
This turning point 
is followed by an early matter-dominated era, when $e_\varphi \gg e_\DS \gg e_\SM$. 
For viable parameter choices, 
$e_\varphi$ vanishes close to the phase transition in the YM sector. 
Although SM fields are sourced both before and after the phase transition, 
the decay of a glueball 
is more likely than the $\mathcal{G} \mathcal{G} \to \textrm{SM}\, \textrm{SM}$ scattering of two dark gluons (c.f.~eqs.~\eqref{GammaGB} and \eqref{NRO_vectors} respectively). 
The consequence is a ``kink'' on the $e_\SM$ line at the beginning 
of the phase transition in the dark sector. 

When the phase transition 
is completed, $e_\DS$ contains both stable and unstable glueballs. 
Their respective fractions vary according to the Boltzmann equations, 
until finally the unstable glueballs decay. 
This happens when 
$H\lesssim \Gamma_{\GB}$. 
The DS consists then of the stable glueballs only, which evolve decoupled from the SM sector, and the standard radiation-dominated era starts.

\subsection{Cosmological and astrophysical  constraints}
Let us now turn to the question of what parameter choices provide a viable scenario.
By viable we mean, in particular, capable of providing the correct relic abundance of sufficiently long-lived dark matter,
while achieving a successful SM reheating. 
In practice, we keep the inflationary dynamics intact and inspect for which dark sector confinement scale, $\LamDS$, and which portal scale, $M$, the following conditions hold 
\begin{align}
&\tau_\DM = 1/\Gamma_\DM \gtrsim 10^{26}\,\mathrm{s} \, ,
\label{tau_condition} \\
&T_\SM^{\text{rh}}
\geq 4 \, \text{MeV} \, , 
\label{SM_condition}
\\
&\Omega_\DM\,h^2 
\leq 0.1200 \pm 0.0012 \, .
\label{DM_condition} 
\end{align}
Eq.~\eqref{tau_condition} corresponds to an approximate bound coming from 
indirect detection experiments, widely used in literature \cite{Eichler:1989br,Ibarra:2013cra}. 
This constraint
is expected to become more stringent, up to $\tau _\DM \sim 10^{28}\,$s 
\cite{Kalashev:2016cre,Cohen:2016uyg}. 
On the other hand, the precise theoretical estimate of $\Gamma_\DM$ crucially depends on the branching ratios of the decaying dark matter to different SM states, and we choose to leave this analysis to future work. 
Moreover, the interpretation of 
eq.~\eqref{tau_condition} in terms of constraints on $\LamDS$ and $M$ is 
affected also by the uncertainties in the non-perturbative input and matching coefficient, $c_8$, as presented in eq.~\eqref{GammaDM}. 
Keeping these opposite effects 
in mind, in this work we resort ourselves to $\tau_\DM  c_8^2 \sim 10^{ 26-\delta}\,$s, $\delta\in[0,1]$.

In eq.~\eqref{SM_condition}, 
$T_{\text{SM}}^{\text{rh}}$ is the SM temperature when its energy density starts dominating, i.e., after the decay of unstable glueballs. 
The Hubble rate at this time satisfies $H \approx \Gamma_\GB$, and, 
approximating $H^2 \approx 8 \pi e_\SM/(3 \mpl^2)$, we find an analytical estimate for the minimal decay rate required by  
eq.~\eqref{SM_condition},
\begin{equation}
    \Gamma_\GB \geq \Gamma_\GB^{\textrm{min}} \approx 5.8 \times 10^{-43} \mpl \, . 
    \label{gamma_min_SM}
\end{equation}
Combining eqs.~\eqref{gamma_min_SM} and \eqref{GammaGB} sets a constraint on a further combination of $\LamDS$ and $M$.
For the results illustrated in fig.~\ref{fig:parameter_space}, we extract $T_{\text{SM}}^{\text{rh}}$ from the numerical solution for $e_\SM$ of the evolution equations, using the SM equation of state provided in \cite{Laine:eos}, 
finding a slightly larger value, $\Gamma_\GB^\textrm{min}\,|_\textrm{num} \approx 1.4\times 10^{-42}\mpl$. Qualitatively, however, the SM reheating temperature indeed depends solely on the value of $\Gamma_\GB$ and not on other combination of the parameters $\LamDS$ and $M$.

Finally, for eq.~\eqref{DM_condition}, the dark matter relic abundance is dictated by the freeze-out dynamics, which requires a numerical solution of the full system, as presented in eqs.~\eqref{eq:EvInf_dec}--\eqref{eq:Ev_1_NR_FO}.

\subsection{Freeze-out dynamics and parameter space}
\label{subsec:solutions}

At the core of the extraction of the dark matter energy density one finds the solution of coupled Boltzmann equations.
We solved the equations for a two-glueballs system, described by
eqs.~\eqref{eq:Ev_6_NR_FO}--\eqref{eq:Ev_1_NR_FO}, and for a three-glueballs system, as in eqs.~\eqref{eq:Ev_6_NR_FO_3gb}-\eqref{eq:Ev_2_NR_FO_3gb}. 
The result shown in fig.~\ref{fig:GB_freeze_out} corresponds to the latter case. 
The observed dynamics is qualitatively the same:
the freeze-out of even states
happens when the $3{\to} 2$ interactions cease to be efficient, 
and a non-zero chemical 
potential develops. 
After this point, the $2 {\to} 2$ reactions maintain 
the value of the chemical potentials 
such that $\nVV/\nVV^\eq \approx \nS/\nS^\eq$. 
The relic abundance of the stable glueballs is set at the point when also the $2{\to} 2$ interactions become inefficient. 

Since the 
$2{\to} 2$ cross section scales as $\LamDS^{-2}$, 
(c.f. 
eq.~\eqref{2to2_cross_averaged}), 
for smaller $\LamDS$, the interaction strength is larger, the freeze-out happens later, and the relic abundance is suppressed.
Quantitatively, the addition of the $2^{++}$ state acts as a doubling of the interaction channels and thus of the cross-sections, which translates into $\nVV$ being half as large at freeze-out completion. 

\begin{figure}
   \centering
    \includegraphics[width=.95\linewidth]{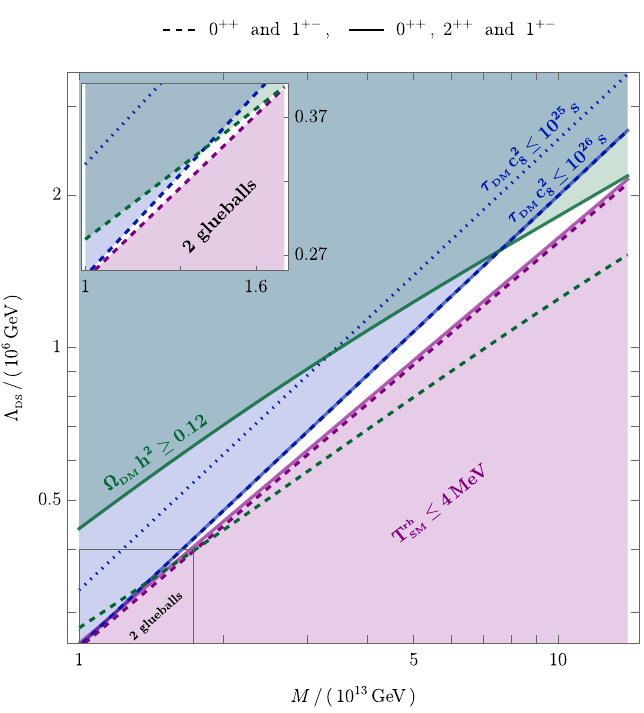}
    \caption{Parameter space of the model in the plane ($M,\LamDS$). The white unshaded triangle is the viable region that complies with the cosmological and astropysical conditions~\eqref{tau_condition}-\eqref{DM_condition}. The solid lines correspond to the three-glueball case described by eqs.~\eqref{eq:Ev_6_NR_FO_3gb}-\eqref{eq:Ev_2_NR_FO_3gb} whereas the dashed lines correspond to the two-glueball one, eqs.~\eqref{eq:Ev_6_NR_FO}-\eqref{eq:Ev_1_NR_FO}. For the latter case, the open window is in the bottom-left corner, shown magnified in the inset (top-left corner).}
    \label{fig:parameter_space}
\end{figure}

The resulting parameter space 
is shown in fig.~\ref{fig:parameter_space}, where the 
viable region 
is unshaded (white).
The interplay of the 
conditions in eqs.~\eqref{tau_condition}--\eqref{DM_condition} is different for the three-glueballs system (solid lines) and for the two-glueballs system (dashed lines). 

The numerical results for $T^{\textrm{rh}}_\SM = 4$ MeV (purple) confirm the estimate in eq.~\eqref{gamma_min_SM}: a too small $\LamDS$ or 
a too large $M$ 
suppress the decay of the even glueballs, 
resulting in an inefficient energy transfer to the visible sector. Adding the $2^{++}$ state has little impact on this line.
On the other hand, the curves depicting $\Omega_\DM\, h^2=0.12$ (green) differ considerably for the 2-glueball and 3-glueball scenarios as expected based on the discussion of the interaction strengths above. 

Our numerical results also confirm that larger $\LamDS$ leads to larger DM relic abundance. Larger $M$, in turn, implies that the decay of the glueballs to SM is less efficient, and radiation domination appears later, leading to more dilution of the DM density.\footnote{Let us note that this dilution also explains why we do not recover a linear dependence $\Omega_\DM h^2 \propto \Lambda_\DS$, as in the usual case, where the dark matter is constituted by the lightest $0^{++}$ glueball state~\cite{Carlson:1992fn,Halverson:2016nfq, Carenza:2022pjd,Carenza:2023eua}.} 
Finally, exclusion limits from indirect detection (blue), c.f.~eq.~\eqref{tau_condition}, are identical for both cases , but affected by the systematic uncertainty discussed above (see the difference between the dotted and solid lines showing the possibility of enlarging the white region).

Let us stress the large impact of adding the next-to-leading glueball contribution to the dark energy density $e_\DS$. 
A realistic determination of the parameter space must therefore take into account higher excitations when the system falls out of equilibrium. 
On the other hand, we expect that adding even heavier glueballs will have less dramatic effect due to more significant Boltzmann suppression of their number densities.
We remark that $\mathcal{O}(1)$ uncertainties are ubiquitous in the modeling of glueball dynamics, which range from the complexity of the dark phase transition to the proper inclusion of glueball interactions. 

\section{Conclusions and outlook}
\label{sec_discussion}
We have presented a model providing successful inflation and reheating of SM fields and, at the same time, a viable dark matter candidate with correct relic abundance. 
In summary, the thermal history that is expected to occur in our framework can be summarized 
as follows. The inflaton decays into non-Abelian dark vectors that 
form a thermal bath of dark vectors at high temperatures. Already at 
this stage, 
portal interactions between the dark and the visible sectors are active, sourcing a non-negligible visible bath. 
After inflation, the expansion of the universe induces a cooling of the dark plasma that undergoes a phase transition from a deconfined to a confined phase, at a temperature 
$\TDS^c\sim\LamDS$. The interactions among glueballs trigger various 
processes that control the number density of the different dark composite states \cite{Forestell:2016qhc,Soni:2016gzf}. Together with the glueball decays into visible particles, 
$3{\to} 2$ and $2{\to} 2$ scatterings enter a network of evolution equations that ultimately determine the dark matter relic density, as well as the SM reheating temperature.

Our framework is 
relatively general and can be applied to different choices in the dark and inflationary sectors. 
In particular, although the viable region in the parameter space appears small for our particular implementation 
(see fig.~\ref{fig:parameter_space}), 
generalizations of our scenario might enlarge it. 
For example, one may consider different gauge groups, 
e.g.~
SO($N$) with $N\,{\geq}\, 8$~\cite{Juknevich:2009ji,Gross:2020zam,Yamada:2023thl}. 
The decay rate of 
stable glueballs scales then as $\LamDS^{2N-3}/M^{2N-4}$, 
such that the blue line in fig.~\ref{fig:parameter_space} would be shifted upwards correspondingly. 
However, apart from few pioneering studies like~\cite{Lau:2017aom, Teper:2018gaw}, the non-perturbative input on the glueball spectrum and interactions is mostly missing in those theories. 
For these groups, the stable glueball 
would likely be too long-lived to be probed by indirect detection experiments. 
On the other hand, gravitational wave signatures might offer further observational constraints, as discussed in~\cite{Kolesova:2023yfp,Halverson:2020xpg,Huang:2020crf,Morgante:2022zvc,Yamada:2022imq}. In particular, the confinement phase transition in different pure Yang-Mills theories is known to be of first-order \cite{Holland:2003kg,Lucini:2005vg}, for the case at hand leading to a peak frequency in the LISA range, $f_0\sim \mathcal{O}(1)\,$Hz \cite{Huang:2020crf}. Per contra, the spectral amplitude might be 
suppressed by the long-lasting matter-dominated period \cite{Kolesova:2023mno}.

Other generalizations may also consider direct decays of the inflaton 
into SM degrees of freedom, e.g.~via a similar axion-like coupling. This has been considered in the literature of warm inflation, e.g.~\cite{Berghaus:2024zfg}. 
Since $\Lambda_{{\text{\tiny QCD}}}{\ll}\, \LamDS$, the couplings satisfy $\alpha_{{\text{\tiny QCD}}} {<} \alpha$, inducing a corresponding hierarchy in the friction terms. 
Hence, the dark sector could still dominate the total energy density 
for a certain amount of time, 
while the DM relic abundance would be reduced and 
the parameter space enlarged. 
Further freedom is also in the details of inflation itself, as these turn out to barely enter the cosmological observables in fig.~\ref{fig:parameter_space}. In scenarios with a larger friction coefficient, $\Upsilon$, the inflaton decays earlier, and our results are unchanged. On the other hand, for feebler interaction between inflaton and the dark sector, the inflaton might affect the freeze-out dynamics.

Finally,
UV completions of our scenario, leading to the operators in eqs.~\eqref{op6} and \eqref{op8}, 
could be related to other puzzles in particle cosmology as neutrino masses or baryon asymmetry. 

\noindent
\\
\emph{Note Added}: During the revision of our manuscript the study of ref.~\cite{McKeen:2024trt} appeared that
investigates a closely related scenario of dark glueballs.

\section*{Acknowledgements}
We thank Mikko Laine for comments on the manuscript and Torsten Bringmann, Jacopo Ghiglieri, Admir Greljo, Giacomo Landini, Harvey Meyer, Marco Panero and Roman Pasechnik for useful discussions. 
The work of S.B. and S.P. is supported by the Swiss National Science Foundation, under the Ambizione grant 2185783 and under grant 212125 respectively. H.K. is supported by the Research Council of Norway under the FRIPRO Young Research Talent grant no.~335388.

\appendix


\section{Glueball Lagrangian and cross sections}
\label{app1_Glueball_Lag}
Based on naive dimensional
analysis 
\cite{Manohar:1983md,Cohen:1997rt,Luty:1997fk} and large-$N$ scaling \cite{tHooft:1973alw,Witten:1979kh}, we consider the following Lagrangian density for a given glueball $\phi_\GB$ \cite{Boddy:2014yra,Forestell:2016qhc,Forestell:2017wov}
\begin{align}
        \mathcal{L} \, \supset \,& \frac{1}{2} (\partial_\mu \phi_\GB)^2 - \frac{a_2}{2!} m_\GB \phi_\GB^2 - \frac{a_3}{3!} \left( \frac{4 \pi}{N}\right) m_\GB \phi_\GB^3 - \frac{a_4}{4!}  \left( \frac{4 \pi}{N}\right)^2 \phi_\GB^4 \, .
        \label{lag_eff_1}
        \end{align}       
In the first line, the glueball kinetic and mass terms are accompanied by three- and four-glueball self-interactions. 
Terms with higher powers of $\phi_\GB$ within the effective potential approach are considered in other works, e.g.,~\cite{Carenza:2022pjd, Carenza:2023eua}. Such additional interactions 
postpone the freeze-out and reduce the dark matter relic abundance. 
However, they also introduce 
additional unknown coefficients within our approach and we choose not to include them.
The coefficients $a_i$ are expected to be of order unity. 

The glueball interaction terms lack a detailed non-perturbative treatment at the moment. Even if the schematic form of the effective Lagrangian in eq.\eqref{lag_eff_1} may hold for interactions among different glueballs, it is strictly appropriate only for scalar glueball states, such as $0^{++}$. 
For the $3 \to 2$ process of the lightest glueball, $0^{++}$, we improve upon the results of ref.~\cite{Forestell:2016qhc} by computing the diagrams at order $(4 \pi /N)^3$ and find
\begin{equation}
      \langle \sigma_{ 000;00 
      } v^2 \rangle \simeq \frac{1805 \, \pi^5 \, a_3^2}{388962 \, N^6 \, m_1^5} \left( 11 a_3^2 + 7  a_4  \right)^2 \, ,
      \label{sigma3to2}
\end{equation}
at leading order in the velocity expansion. The cross-section in eq.~\eqref{sigma3to2} is about five times smaller than the estimation in refs.~\cite{Forestell:2016qhc,Forestell:2017wov} for $a_i{=}1$, which reads 
$\langle \sigma_{ 000;00 } v^2 \rangle {\simeq} (4 \pi)^3/(N^6 m^5_{0^{++}})$. 

Proceeding with analogous computations for the decays and interactions involving glueballs of higher angular momenta would require a precise formulation of the corresponding effective Lagrangians, which goes beyond the goal of this paper.
In order to align the approximation on the inputs for our equations, we eventually adopt the estimation in terms of the power counting and use \cite{Forestell:2016qhc,Forestell:2017wov}.
The corresponding cross-section for the next-to-lightest C-even glueball, $2^{++}$ is then
\begin{equation}
      \langle \sigma_{ 222;22 
      } v^2 \rangle \simeq  \frac{(4 \pi)^3}{N^6 \, m_{2^{++}}^5} \, .
      \label{sigma3to2_2pp}
\end{equation} 
Also for the $2 {\to} 2$ processes, we simply estimate of the squared amplitude in terms of the power counting of the effective Lagrangian, and we obtain at leading order in the velocity expansion and with $a_i=1$ the expression
\begin{align}
   &\langle \sigma_{a b ; 
   cd} v \rangle \simeq 
   \frac{(4\pi/N)^4}{g_a g_b S_{cd} \, 64 \pi^2  s} 
   \frac{ \lambda^{1/2}\left( 1,m_c^2/s,m_d^2/s \right) }{
   \lambda^{1/2}\left( 1,m_a^2/s,m_b^2/s \right) }\, .
   \label{2to2_cross_averaged}
\end{align}
Here, $\lambda(x,y,z)=(x-y-z)^2-4yz$ is the K\"all\'en function, $g_i =(2 J_i+1)$, $s=(m_a+m_b)^2$, and $S_{cd}$ is the symmetry factor accounting for identical particles in the final state.

\section{Three-glueballs evolution equations}
\label{app2_Glueball_3}
The evolution equations describing the 
freeze-out dynamics of two unstable states, $0^{++}$ and $2^{++}$, and one stable, $1^{+-}$, read
\begin{align}
    \dot{n}_{1^{+-}}  + 3H \nVV 
    =&\, 
    B_{\varphi}^{1^{+-}} \Upsilon\frac{e_\varphi}{m_{1^{+-}}}  
    -\langle \sigma_{11;00} v \rangle 
        \left[ \nVV^2 - \left( \frac{\nS}{\nS^\eq} \right)^2  (\nVV^\eq)^2 \right]  \nonumber\\
    &-\langle \sigma_{11;22} v \rangle 
        \left[ \nVV^2 - \left( \frac{\nSS}{\nSS^\eq} \right)^2  (\nVV^\eq)^2 \right] \nonumber\\    
    &-\langle \sigma_{11;02} v \rangle 
        \left[ \nVV^2 -  \frac{\nS\nSS}{\nS^\eq\nSS^\eq} (\nVV^\eq)^2 \right]  \,, \label{eq:Ev_6_NR_FO_3gb}\\ 
    \dot{n}_{0^{++}} + 3H \nS   
    =&\, B_{\varphi}^{0^{++}} \Upsilon\frac{e_\varphi}{m_{0^{++}}}   - \nS \Gamma_{0^{++}} + \nSS \Gamma_{2^{++}}
   \nonumber \\
    &+\langle \sigma_{11;00} v \rangle \left[ \nVV^2 - \left( \frac{\nS}{\nS^\eq} \right)^2  (\nVV^\eq)^2 \right]  \nonumber\\
    &- \langle \sigma_{000;00} v^2 \rangle \,\nS^2(\nS- \nS^\eq ) \nonumber \\
    &+ \langle \sigma_{22;00} v \rangle 
        \left[ \nSS^2 - \left( \frac{\nS}{\nS^\eq} \right)^2  (\nSS^\eq)^2 \right]  \nonumber\\
    &+ \frac{ \langle \sigma_{11;02} v \rangle}{2} 
        \left[ \nVV^2 -  \frac{\nS\nSS}{\nS^\eq\nSS^\eq} (\nVV^\eq)^2 \right]  \nonumber\\
    &+ \left[ \nS\langle \sigma_{20;00} v \rangle + \frac{\nSS\langle \sigma_{22;20} v \rangle}{2} \right.\nonumber\\
    &\hphantom{+[} \left. \vphantom{\frac{1}{2}} + \nVV\langle \sigma_{12;10} v \rangle \right] 
         \left( \nSS -  \frac{\nSS^\eq}{\nS^\eq} \nS \right)  
    \, ,\label{eq:Ev_1_NR_FO_3gb}  \\
     \dot{n}_{2^{++}} + 3H \nSS   
    =&\, B_{\varphi}^{2^{++}} \Upsilon\frac{e_\varphi}{m_{2^{++}}}   - \nSS \Gamma_{2^{++}}
   \nonumber \\
    &+\langle \sigma_{11;22} v \rangle \left[ \nVV^2 - \left( \frac{\nSS}{\nSS^\eq} \right)^2  (\nVV^\eq)^2 \right]  \nonumber\\
    &- \langle \sigma_{222;22} v^2 \rangle \,\nSS^2(\nSS- \nSS^\eq ) \nonumber \\
    &- \langle \sigma_{22;00} v \rangle 
        \left[ \nSS^2 - \left( \frac{\nS}{\nS^\eq} \right)^2  (\nSS^\eq)^2 \right]  \nonumber\\
    &+ \frac{ \langle \sigma_{11;02} v \rangle}{2} 
        \left[ \nVV^2 -  \frac{\nS\nSS}{\nS^\eq\nSS^\eq} (\nVV^\eq)^2 \right]  \nonumber\\
    &- \left[ \nS\langle \sigma_{20;00} v \rangle + \nSS\frac{\langle \sigma_{22;20} v \rangle}{2} \right.\nonumber\\
    &\hphantom{+[} \left.\vphantom{\frac{1}{2}} + \nVV\langle \sigma_{12;10} v \rangle \right] 
         \left( \nSS -  \frac{\nSS^\eq}{\nS^\eq} \nS \right)  
    \, .\label{eq:Ev_2_NR_FO_3gb} 
\end{align}
The lightest 
state, $0^{++}$, 
decays 
to the SM 
via $\Gamma_{0^{++}}=\Gamma_\GB$ in eq.~\eqref{GammaGB}.
Since the next-to-lightest state, $2^{++}$, has spin two, it decays through different operators. 
In particular, the direct decay to SM fields 
in analogy to $\Gamma_\GB$ implies a dimension-eight operator.
However, a dimension-six contribution is allowed for our setup 
via the process $2^{++}{\to}0^{++}h$ 
\cite{Juknevich:2009gg}. 
Its decay rate is 
\begin{align}
    \langle \Gamma_{2^{++}} \rangle
    \simeq 
    \frac{ c_6^2 }{480 \pi} m_{2^{++}}^3 \left(\frac{ \alpha M_S^{ 2^{++},0^{++} } }{M^2}\right)^2 \left( 1-\frac{m_{0^{++}}}{m_{2^{++}}}\right)^5 \left.\frac{K_1(z)}{K_2(z)}\right|_{z=m_{2^{++}/T_{\scalebox{.45}{DS}}}} \,,
\end{align}
with the
matrix element $M_S^{ 2^{++},\,0^{++} }\!{\approx} \,0.03\, m_{0^{++}}$~\cite{Juknevich:2009ji}.

\bibliographystyle{elsarticle-num} 
\bibliography{paper_biblio} 






\end{document}